%% file: main.tex
\pdfoutput=1
\documentclass[12pt,a4paper]{article}
\usepackage{ifthen} % for conditional statements
\newboolean{pdflatex}
\setboolean{pdflatex}{true} % False for eps figures 

\newboolean{articletitles}
\setboolean{articletitles}{true} % False removes titles in references

\newboolean{uprightparticles}
\setboolean{uprightparticles}{false} %True for upright particle symbols

\newboolean{inbibliography}
\setboolean{inbibliography}{false} %True once you enter the bibliography

\input{preamble}

\begin{document}
\newcommand{\cfft}{C$_{4}$F$_{10}$ }
\newcommand{\cff}{CF$_{4}$ }
\renewcommand{\floatpagefraction}{.8}

%%%%%%%%%%%%%%%%%%%%%%%%%
%%%%% Title     %%%%%%%%%
%%%%%%%%%%%%%%%%%%%%%%%%%
\renewcommand{\thefootnote}{\fnsymbol{footnote}}
\setcounter{footnote}{1}

% %%%%%%% CHOOSE TITLE PAGE--------
%\onecolumn
\input{title-LHCb-INT}
%\twocolumn
% %%%%%%%%%%%%% ---------

\renewcommand{\thefootnote}{\arabic{footnote}}
\setcounter{footnote}{0}

%%%%%%%%%%%%%%%%%%%%%%%%%%%%%%%%
%%%%%  Table of Content   %%%%%%
%%%%%%%%%%%%%%%%%%%%%%%%%%%%%%%%
%%%% Uncomment next 2 lines if desired
%\tableofcontents
%\cleardoublepage

%%%%%%%%%%%%%%%%%%%%%%%%%
%%%%% Main text %%%%%%%%%
%%%%%%%%%%%%%%%%%%%%%%%%%

\pagestyle{plain} % restore page numbers for the main text
\setcounter{page}{1}
\pagenumbering{arabic}

%% Uncomment during review phase. 
%% Comment before a final submission.
%\linenumbers

\input{mypaper}

\addcontentsline{toc}{section}{References}
\setboolean{inbibliography}{true}
\bibliographystyle{LHCb}
\bibliography{main}

\end{document}

%% file: preamble.tex
\usepackage[top=1in, bottom=1.25in, left=1in, right=1in]{geometry}

\renewcommand{\floatpagefraction}{0.99}

\usepackage{microtype}
\usepackage{lineno}  % for line numbering during review
\usepackage{xspace} % To avoid problems with missing or double spaces after
\usepackage{caption} %these three command get the figure and table captions automatically small

\usepackage{graphicx}  % to include figures (can also use other packages)
\usepackage{color}
\usepackage{colortbl}
\graphicspath{{./figs/}} % Make Latex search fig subdir for figures

\usepackage{amsmath} % Adds a large collection of math symbols
\usepackage{amssymb}
\usepackage{amsfonts}
\usepackage{upgreek} % Adds in support for greek letters in roman typeset

\newcommand*\patchAmsMathEnvironmentForLineno[1]{%
\expandafter\let\csname old#1\expandafter\endcsname\csname #1\endcsname
\expandafter\let\csname oldend#1\expandafter\endcsname\csname
end#1\endcsname
 \renewenvironment{#1}%
   {\linenomath\csname old#1\endcsname}%
   {\csname oldend#1\endcsname\endlinenomath}%
}
\newcommand*\patchBothAmsMathEnvironmentsForLineno[1]{%
  \patchAmsMathEnvironmentForLineno{#1}%
  \patchAmsMathEnvironmentForLineno{#1*}%
}
\AtBeginDocument{%
\patchBothAmsMathEnvironmentsForLineno{equation}%
\patchBothAmsMathEnvironmentsForLineno{align}%
\patchBothAmsMathEnvironmentsForLineno{flalign}%
\patchBothAmsMathEnvironmentsForLineno{alignat}%
\patchBothAmsMathEnvironmentsForLineno{gather}%
\patchBothAmsMathEnvironmentsForLineno{multline}%
\patchBothAmsMathEnvironmentsForLineno{eqnarray}%
}

\usepackage[bookmarks=false]{hyperref}    % Hyperlinks in references
\input{lhcb-symbols-def} % Add in the predefined LHCb symbols

\usepackage{cite} % Allows for ranges in citations
\usepackage{mciteplus}

%% file: lhcb-symbols-def.tex
\usepackage{xspace} 
\usepackage{upgreek}

\def\MagUp {\mbox{\em Mag\kern -0.05em Up}\xspace}

\ifthenelse{\boolean{uprightparticles}}%
{

 \def\PDelta      {\ensuremath{\Delta}\xspace}                 
 \def\PXi      {\ensuremath{\Xi}\xspace}                 
 \def\PLambda      {\ensuremath{\Lambda}\xspace}                 
 \def\PSigma      {\ensuremath{\Sigma}\xspace}                 
 \def\POmega      {\ensuremath{\Omega}\xspace}                 
 \def\PUpsilon      {\ensuremath{\Upsilon}\xspace}                 
 
 \def\PB      {\ensuremath{\mathrm{B}}\xspace}                 
                  
 \def\PD      {\ensuremath{\mathrm{D}}\xspace}

 \def\PK      {\ensuremath{\mathrm{K}}\xspace}

 \def\Pi      {\ensuremath{\mathrm{i}}\xspace}

}
{

 \mathchardef\PDelta="7101
 \mathchardef\PXi="7104
 \mathchardef\PLambda="7103
 \mathchardef\PSigma="7106
 \mathchardef\POmega="710A
 \mathchardef\PUpsilon="7107
                  
 \def\PB      {\ensuremath{B}\xspace}                 
                  
 \def\PD      {\ensuremath{D}\xspace}

 \def\PK      {\ensuremath{K}\xspace}

 \def\Pi      {\ensuremath{i}\xspace}

}

\makeatletter
\ifcase \@ptsize \relax% 10pt
  \newcommand{\miniscule}{\@setfontsize\miniscule{4}{5}}% \tiny: 5/6
\or% 11pt
  \newcommand{\miniscule}{\@setfontsize\miniscule{5}{6}}% \tiny: 6/7
\or% 12pt
  \newcommand{\miniscule}{\@setfontsize\miniscule{5}{6}}% \tiny: 6/7
\fi
\makeatother

\DeclareRobustCommand{\optbar}[1]{\shortstack{{\miniscule (\rule[.5ex]{1.25em}{.18mm})}
  \\ [-.7ex] $#1$}}

  \def\Kbar    {{\kern 0.2em\overline{\kern -0.2em \PK}{}}\xspace}

\def\KorKbar    {\kern 0.18em\optbar{\kern -0.18em K}{}\xspace}

  \def\Dbar    {{\kern 0.2em\overline{\kern -0.2em \PD}{}}\xspace}

\def\DorDbar    {\kern 0.18em\optbar{\kern -0.18em D}{}\xspace}

\def\Bbar    {{\ensuremath{\kern 0.18em\overline{\kern -0.18em \PB}{}}}\xspace}

\def\BorBbar    {\kern 0.18em\optbar{\kern -0.18em B}{}\xspace}

  \def\Y#1S{\ensuremath{\PUpsilon{(#1S)}}\xspace}% no space before {...}!

\def\Lbar        {{\ensuremath{\kern 0.1em\overline{\kern -0.1em\PLambda}}}\xspace}
\def\LorLbar    {\kern 0.18em\optbar{\kern -0.18em \PLambda}{}\xspace}

\def\AT#1     {\ensuremath{A_{\mathrm{T}}^{#1}}\xspace}           % 2

\def\C#1      {\ensuremath{\mathcal{C}_{#1}}\xspace}                       % 9
\def\Cp#1     {\ensuremath{\mathcal{C}_{#1}^{'}}\xspace}                    % 7
\def\Ceff#1   {\ensuremath{\mathcal{C}_{#1}^{\mathrm{(eff)}}}\xspace}        % 9  
\def\Cpeff#1  {\ensuremath{\mathcal{C}_{#1}^{'\mathrm{(eff)}}}\xspace}       % 7
\def\Ope#1    {\ensuremath{\mathcal{O}_{#1}}\xspace}                       % 2
\def\Opep#1   {\ensuremath{\mathcal{O}_{#1}^{'}}\xspace}                    % 7

\newcommand{\tev}{\ifthenelse{\boolean{inbibliography}}{\ensuremath{~T\kern -0.05em eV}}{\ensuremath{\mathrm{\,Te\kern -0.1em V}}}\xspace}
\newcommand{\gev}{\ensuremath{\mathrm{\,Ge\kern -0.1em V}}\xspace}
\newcommand{\mev}{\ensuremath{\mathrm{\,Me\kern -0.1em V}}\xspace}
\newcommand{\kev}{\ensuremath{\mathrm{\,ke\kern -0.1em V}}\xspace}
\newcommand{\ev}{\ensuremath{\mathrm{\,e\kern -0.1em V}}\xspace}
\newcommand{\gevc}{\ensuremath{{\mathrm{\,Ge\kern -0.1em V\!/}c}}\xspace}
\newcommand{\mevc}{\ensuremath{{\mathrm{\,Me\kern -0.1em V\!/}c}}\xspace}
\newcommand{\gevcc}{\ensuremath{{\mathrm{\,Ge\kern -0.1em V\!/}c^2}}\xspace}
\newcommand{\gevgevcccc}{\ensuremath{{\mathrm{\,Ge\kern -0.1em V^2\!/}c^4}}\xspace}
\newcommand{\mevcc}{\ensuremath{{\mathrm{\,Me\kern -0.1em V\!/}c^2}}\xspace}

\def\gsim{{~\raise.15em\hbox{$>$}\kern-.85em
          \lower.35em\hbox{$\sim$}~}\xspace}
\def\lsim{{~\raise.15em\hbox{$<$}\kern-.85em
          \lower.35em\hbox{$\sim$}~}\xspace}

\def\tell1  {TELL1\xspace}
\def\ukl1   {UKL1\xspace}

%% file: title-LHCb-INT.tex
% $Id: title-LHCb-INT.tex  $
% ===============================================================================
% Purpose: LHCb-INT Note title page template
% Author: P. Koppenburg
% Created on: 2015-05-18
% ===============================================================================

%%%%%%%%%%%%%%%%%%%%%%%%%
%%%%%  TITLE PAGE  %%%%%%
%%%%%%%%%%%%%%%%%%%%%%%%%
\begin{titlepage}

% Header ---------------------------------------------------
\vspace*{-1.5cm}

\noindent
\begin{tabular*}{\linewidth}{lc@{\extracolsep{\fill}}r@{\extracolsep{0pt}}}
\ifthenelse{\boolean{pdflatex}}% Logo format choice
{\vspace*{-2.7cm}\mbox{\!\!\!\includegraphics[width=.14\textwidth]{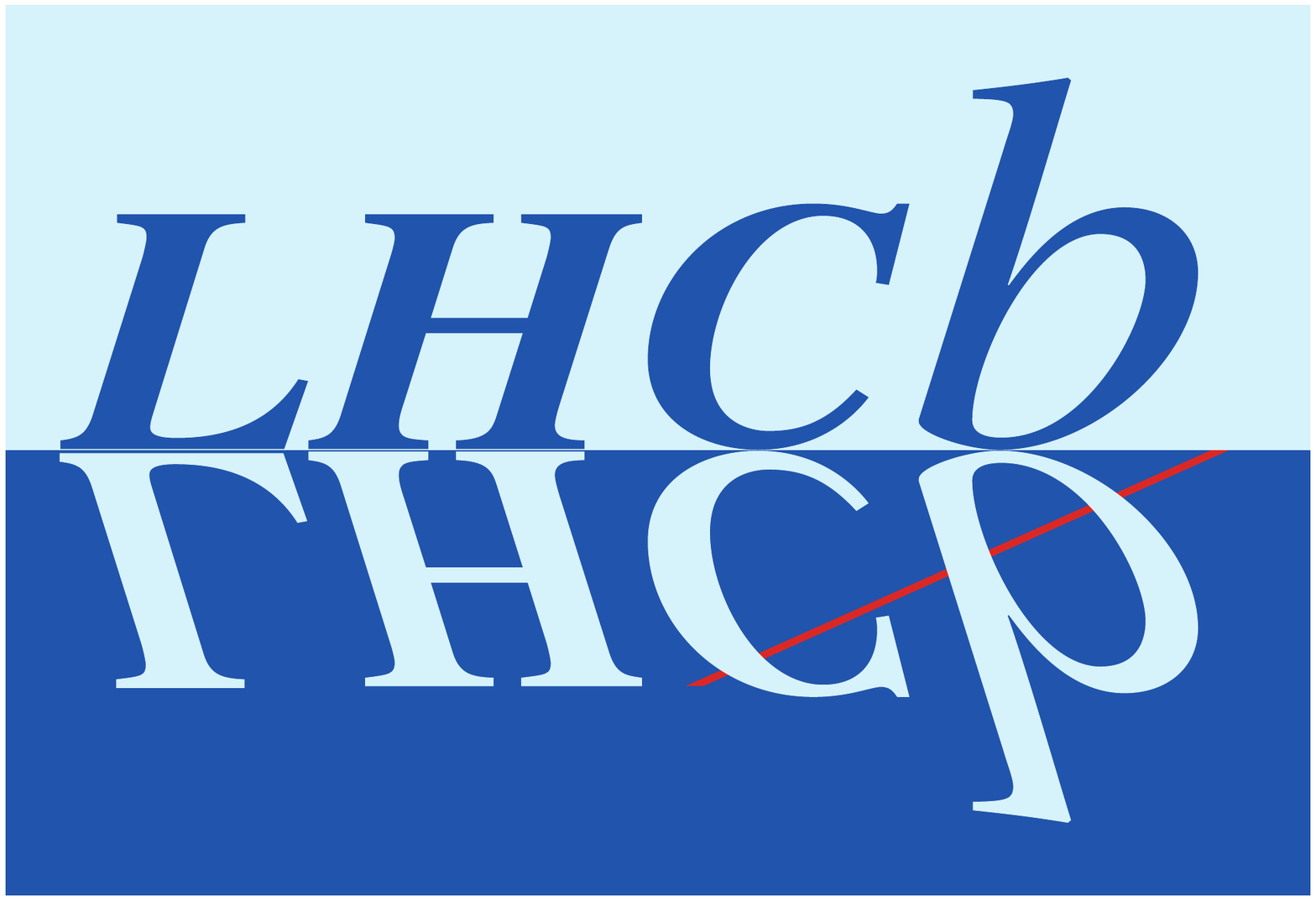}} & &}%
{\vspace*{-1.2cm}\mbox{\!\!\!\includegraphics[width=.12\textwidth]{lhcb-logo.eps}} & &}
 \\
 & & LHCb-PUB-2017-012 \\  % ID 
% & & \today \\ % Date - Can also hardwire e.g.: 23 March 2010
 & & 17 March 2017 \\
 & & \\
\hline
\end{tabular*}

\vspace*{4.0cm}

% Title --------------------------------------------------
{\normalfont\bfseries\boldmath\huge
\begin{center}
  Performance of the LHCb RICH detectors during the LHC Run II
\end{center}
}

\vspace*{2.0cm}

% Authors -------------------------------------------------
\begin{center}
A.~Papanestis$^1$, C.~D'Ambrosio$^2$
\bigskip\\
{\normalfont\itshape\footnotesize
$ ^1$STFC - Rutherford Appleton Laboratory, Didcot, United Kingdom\\
$ ^2$CERN, Geneva, Switzerland\\
}
\vspace*{0.6cm}
on behalf of the LHCb RICH Collaboration
\end{center}
\vspace{\fill}

% Abstract -----------------------------------------------
\begin{abstract}
  \noindent
The LHCb RICH system provides hadron identification over a wide momentum range 
(2--100 GeV/$c$). This detector system is key to LHCb's  precision flavour 
physics programme, which has unique sensitivity to physics beyond 
the standard model. This paper reports on the performance of the LHCb RICH 
in Run II, following significant changes in the detector and operating 
conditions. 
The changes include the refurbishment of significant number of photon 
detectors, assembled
using new vacuum technologies, and the removal of the aerogel radiator. 
The start of Run II of the LHC saw the beam energy
increase to 6.5~TeV per beam and a new trigger strategy for LHCb with full
online detector calibration.
The RICH information has also been made available for all trigger streams
in the High Level Trigger for the first time.
\end{abstract}

\vspace*{2.0cm}
\vspace{\fill}

\end{titlepage}

\pagestyle{empty}  % no page number for the title 

%%%%%%%%%%%%%%%%%%%%%%%%%%%%%%%%
%%%%%  EOD OF TITLE PAGE  %%%%%%
%%%%%%%%%%%%%%%%%%%%%%%%%%%%%%%%

%  empty page follows the title page ----
\newpage
\setcounter{page}{2}
\mbox{~}

\cleardoublepage

%% file: mypaper.tex
\section{Introduction}
\begin{figure}[b]
\centering
\includegraphics[width=0.45\linewidth]{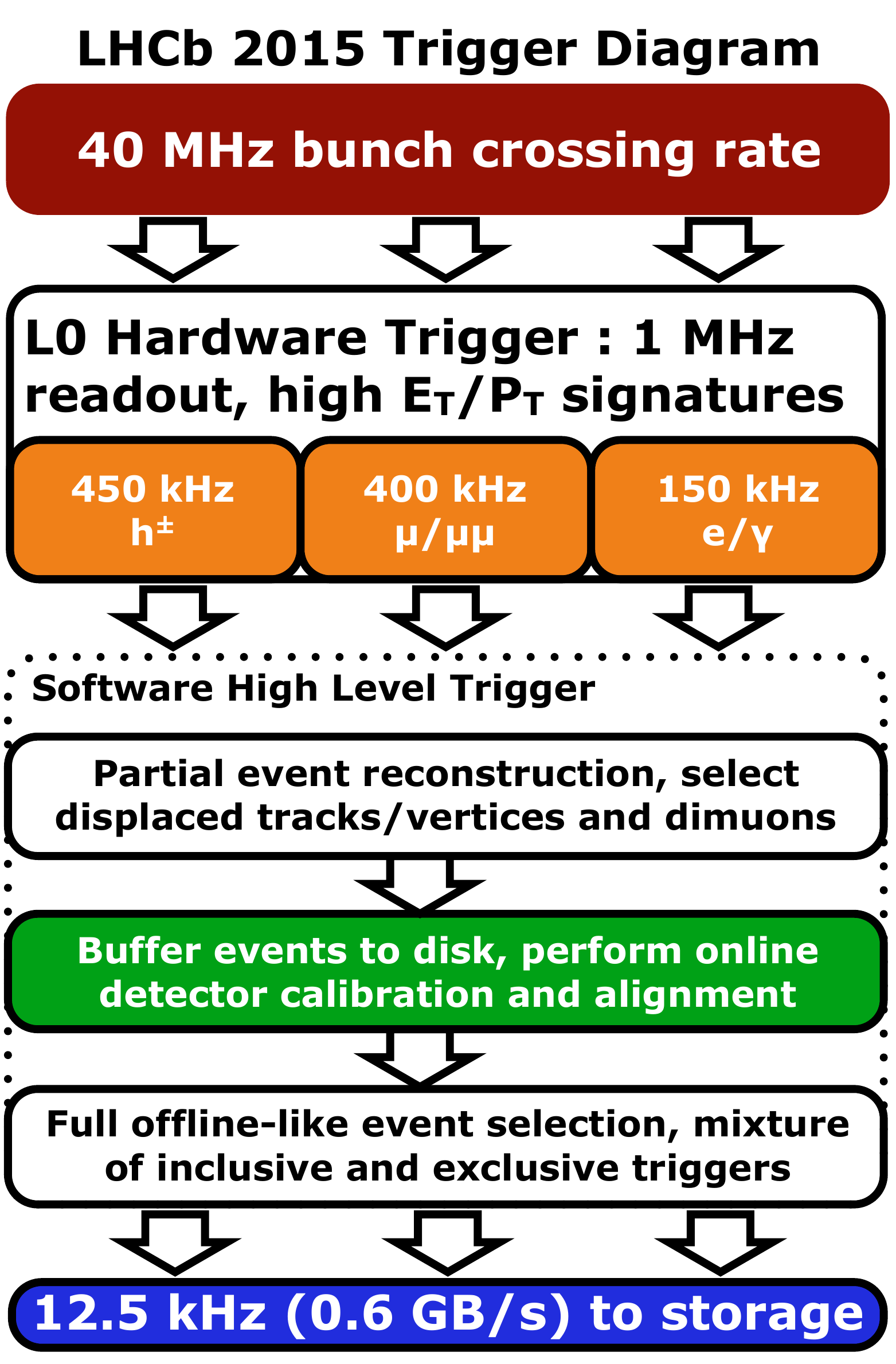}
\caption{The LHCb Run II trigger scheme illustrating the importance of the 
detector alignment and calibration before the HLT2 step.}
\label{fig:trigger}
\end{figure}

After the start of Run II in 2015 at the LHC\cite{Evans:2009zzc} the LHCb 
experiment\cite{LHCb-DP-2014-002} 
has been able to study
the properties and decays of mesons containing $b$ and $c$ quarks 
produced at a $pp$ centre-of-mass energy
of 13~TeV. The crucial task of hadron identification is performed
by a Ring Imaging Cherenkov (RICH) system using two RICH detectors 
with gas radiators\cite{LHCb-DP-2012-003, Papanestis:2014tya}.
For Run II LHCb has been able to collect data with a much higher efficiency
thanks to a new trigger strategy. The Level 0 hardware trigger
remains the same as in Run I (2009--2013), however, the High Level software Trigger (HLT) has
been split into two parts. HLT1 is a fast track-based trigger that runs online
with an output rate of $\sim$100 kHz.
The data are buffered on local disks, and then read by the HLT2, after 
the sub-detector systems have been calibrated and aligned (see 
Figure~\ref{fig:trigger}). 
This results in almost offline-quality datasets straight from the LHCb trigger.
For this strategy to work it is required that the same
algorithms are used online and offline. This increases the need for speed
optimisation in the event reconstruction. These changes had a big impact 
on all subsystems, including the RICH\cite{Jibo}. With the new trigger strategy
the RICH particle identification (ID) information is used in HLT2 for all trigger streams.
Some physics analyses take advantage of this and can be done without the 
extra step of offline reprocessing of the events.

\section{Changes in the RICH System for Run II}
\label{changes}
\begin{figure}
\centering
\includegraphics[width=\linewidth]{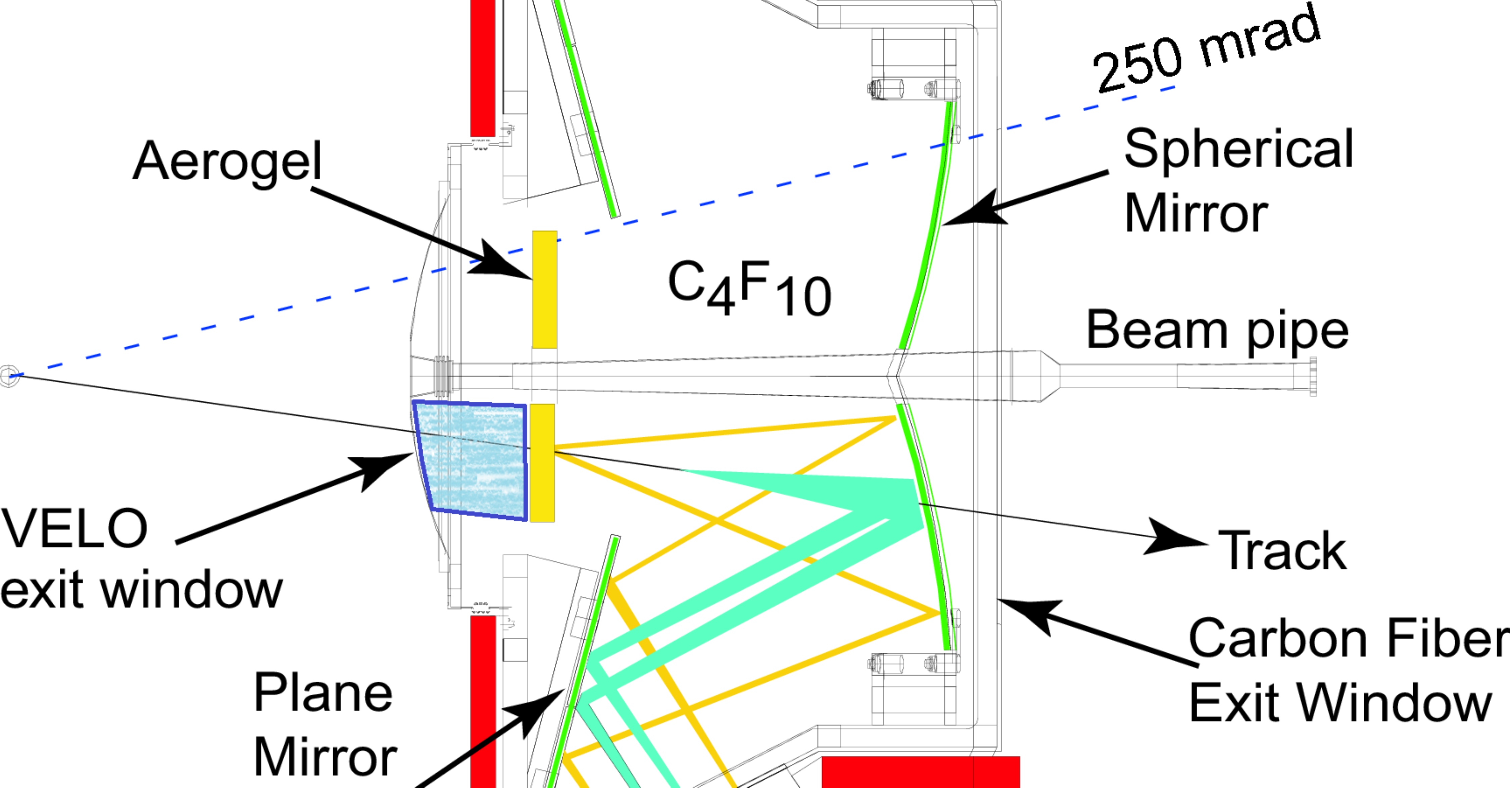}
\caption{A section of the RICH\,1 schematic showing the gas volume from 
the side.
While the aerogel radiator was in place the Cherenkov photons generated in 
the shaded blue area filled with C$_{4}$F$_{10}$ would be blocked. With
the removal of the box these photons are free to reach the photon detectors. }
\label{fig:r1}
\end{figure}

The LHCb RICH system consists of two RICH detectors: RICH\,1 located upstream
of the LHCb magnet and close to the interaction point and RICH\,2 located
downstream of the tracking system and before the calorimeter. Both detectors
have gas radiators (\cfft and \cff respectively) while RICH\,1 also
used aerogel in Run I.
The RICH detectors were re-optimised following Run I taking into account the 
new requirements for the event reconstruction and the need to run the 
same algorithms online and offline. 
The aerogel radiator was removed as its ability to provide particle ID for
particles below the \cfft Cherenkov threshold for kaons is compromised 
by the total number of photons in RICH\,1 in such a high track multiplicity
environment. At the same time its removal allowed the
full use of the gas radiator that is located between the 
RICH\,1 entrance window
and the aerogel (Figure~\ref{fig:r1}, shaded blue area). 
Removing it also contributed 
significantly to the speed of the RICH reconstruction as it reduced 
by more than half the number of photon candidates (combinations
of photon-detector hits with tracks) for which a Cherenkov angle 
is calculated.

The operation of the LHCb RICH Hybrid Photon Detectors (HPDs) and the observed
vacuum degradation on a small number of them has been documented
in the past\cite{Eisenhardt:2014qqa}. 
With no need for HPD refurbishment between Run I and II, the manufacturer
improved the refurbishing process resulting in HPDs with extremely low
ion feedback (i.e.\ very good vacuum quality) and no sign of vacuum 
degradation.

There are also small improvements in the composition of the gas radiators.
In RICH\,1 a liquefying stage was added in the gas re-circulation that 
allows the purification of the \cfft gas (mainly the removal of nitrogen) when
needed. In RICH\,2 the average amount of CO$_{2}$ was increased to about
10\% in order to reduce the \cff scintillation.

\subsection{Cherenkov angle resolution}
\begin{figure}
\centering
\includegraphics[width=0.85\linewidth]{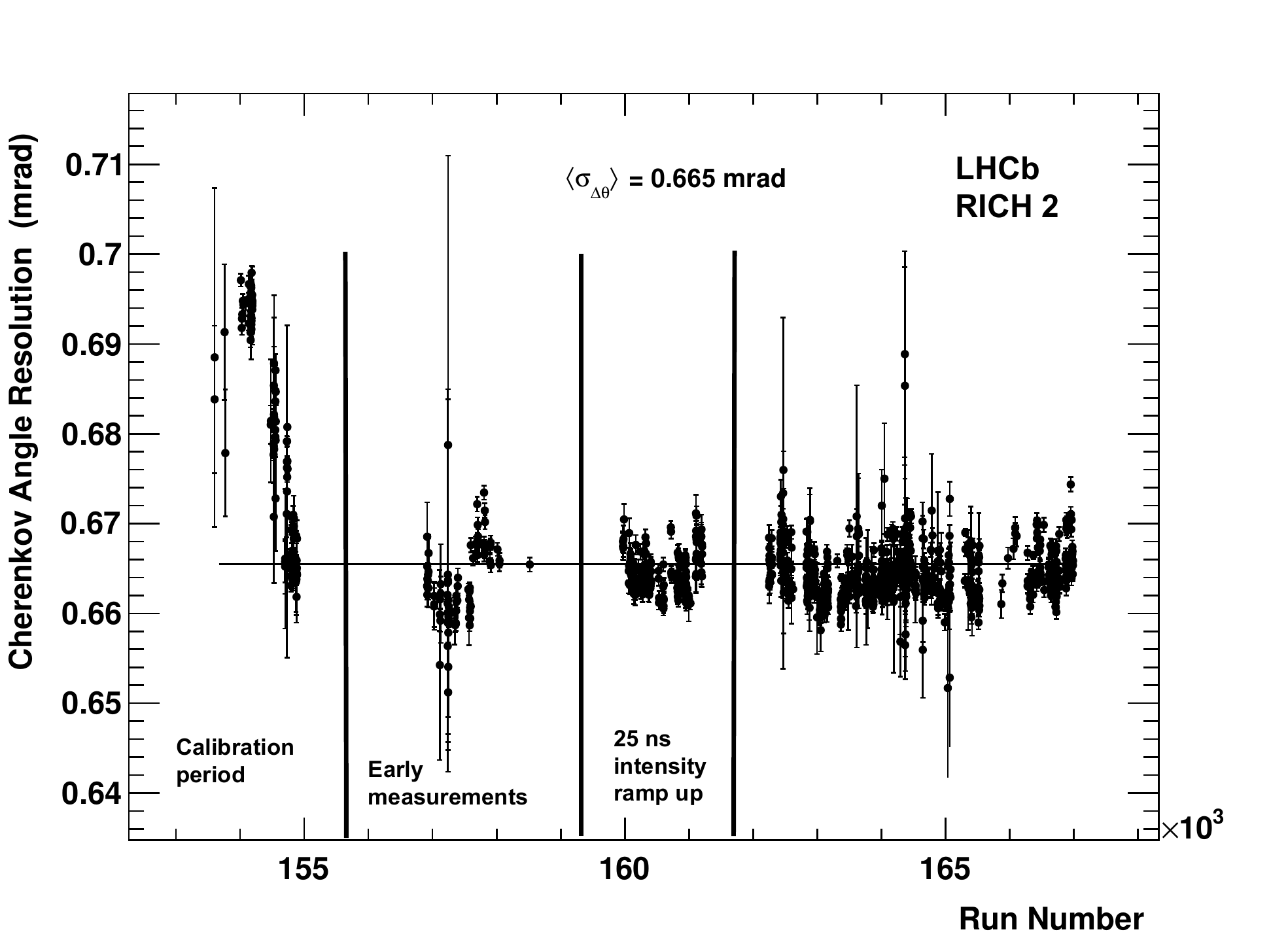}
\caption{RICH\,2 single photon resolution for the summer and early autumn
of 2015. Every point represents the resolution in one run with a 
maximum duration of one hour. The horizontal line is the average of all
the runs after 155000.}
\label{fig:R2res}
\end{figure}

The changes mentioned above were not expected to affect the Cherenkov
angle resolution. Indeed the single photon resolution has remained 
unchanged between Run I and Run II. The average value for RICH\,1 is
1.65~mrad and for RICH\,2 0.67~mrad. Figure~\ref{fig:R2res} shows
a plot with a resolution value for every physics run in the summer 
and early autumn of 2015 for RICH\,2. The maximum duration for a run 
is one hour. After the initial alignment of the mirrors (runs before 
155000), the resolution is very stable, giving
consistent performance throughout the year.

\section{RICH Particle ID Calibration}
\label{calib}

\begin{table}
\centering
\caption{Decays with well defined topology which can be efficiently 
identified without RICH information}
\vspace{2mm}
\begin{tabular}{ l  l  l }
  Species    & Low $p$ and $p_{t}$   & High $p$ and $p_{t}$ \\		
\hline	
  $\pi^{\pm}$ & $K^{0}_{s} \rightarrow \pi^{+} \pi ^{-}$ & 
      $D^{*} \rightarrow D^{0}(K^{-}\pi^{+})\pi^{+}$ \\
  $K^{\pm}$   & $D^{+}_{s} \rightarrow K^{+} K^{-} \pi ^{+}$ &
      $D^{*} \rightarrow D^{0}(K^{-}\pi^{+})\pi^{+}$ \\
  $p^{\pm}$   & $\Lambda^{0} \rightarrow p \pi^{-}$ &  
      $\Lambda^{0} \rightarrow p\pi^{-}, \Lambda^{+}_{c} \rightarrow p K^{-}\pi^{+}$\\
%\hline  
\end{tabular}
\label{tab:particles}
\end{table}

\begin{figure}
\centering
\includegraphics[width=\linewidth]{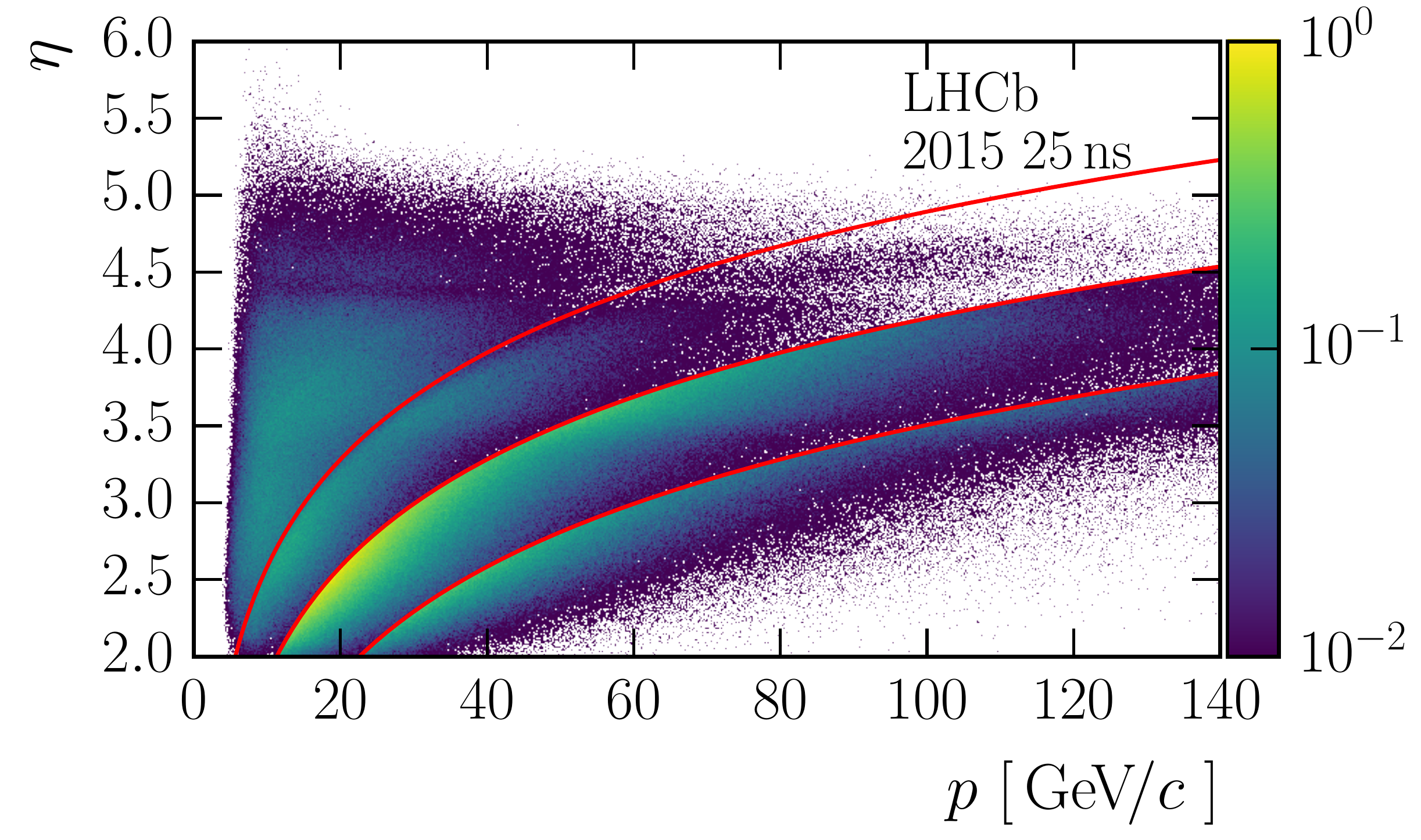}
\vspace{-7mm}
\caption{The kinematic range in momentum $p$ and pseudorapidity $\eta$ of 
the calibration
samples for protons. The red lines correspond to $p_{T}$ thresholds of 1.5, 3.0 
and 6.0 GeV/$c$. The units on the colour bar scale are arbitrary.}
\label{fig:range}
\end{figure}

The performance of the RICH system in particle identification is obtained
from data. It is possible to obtain pure particle samples of known species
(pions, kaons and protons) from decays with well defined topologies
(seen in Table~\ref{tab:particles}) without using the RICH 
\cite{Lupton:2134057} system.
Comparison of the known particles from the calibration samples with
the RICH identification quantifies how well the 
RICH system performs. Care must be taken to ensure that the calibration samples
share similar ranges in momentum and angular distributions
with the physics analysis samples. In Run II
there has been an increase in the range of momentum and pseudorapidity of the
calibration samples to better match the physics events of interest 
(Figure~\ref{fig:range} for protons).

\section{Particle Identification Performance}
\begin{figure}
\centering
\includegraphics[width=\linewidth]{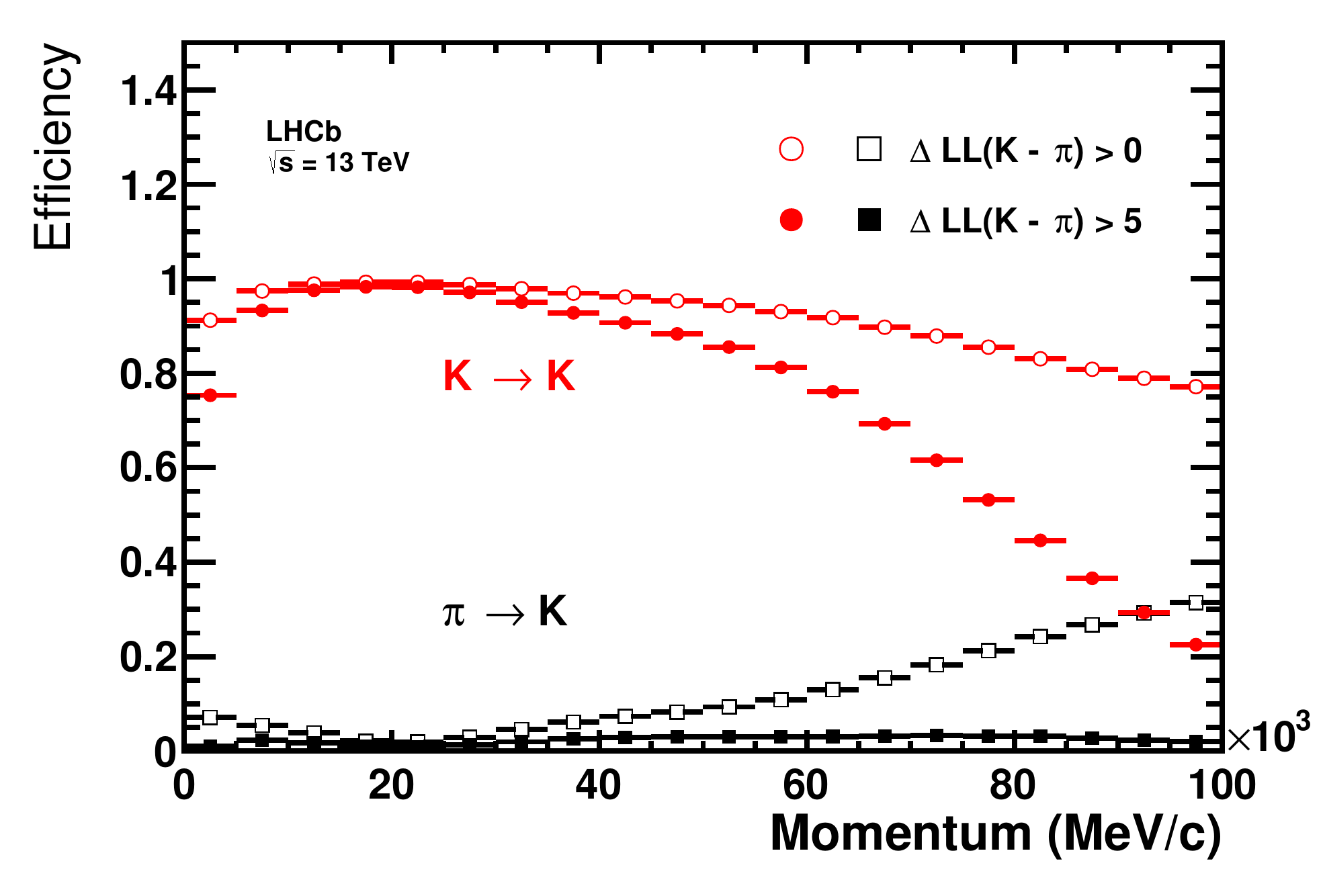}
\includegraphics[width=\linewidth]{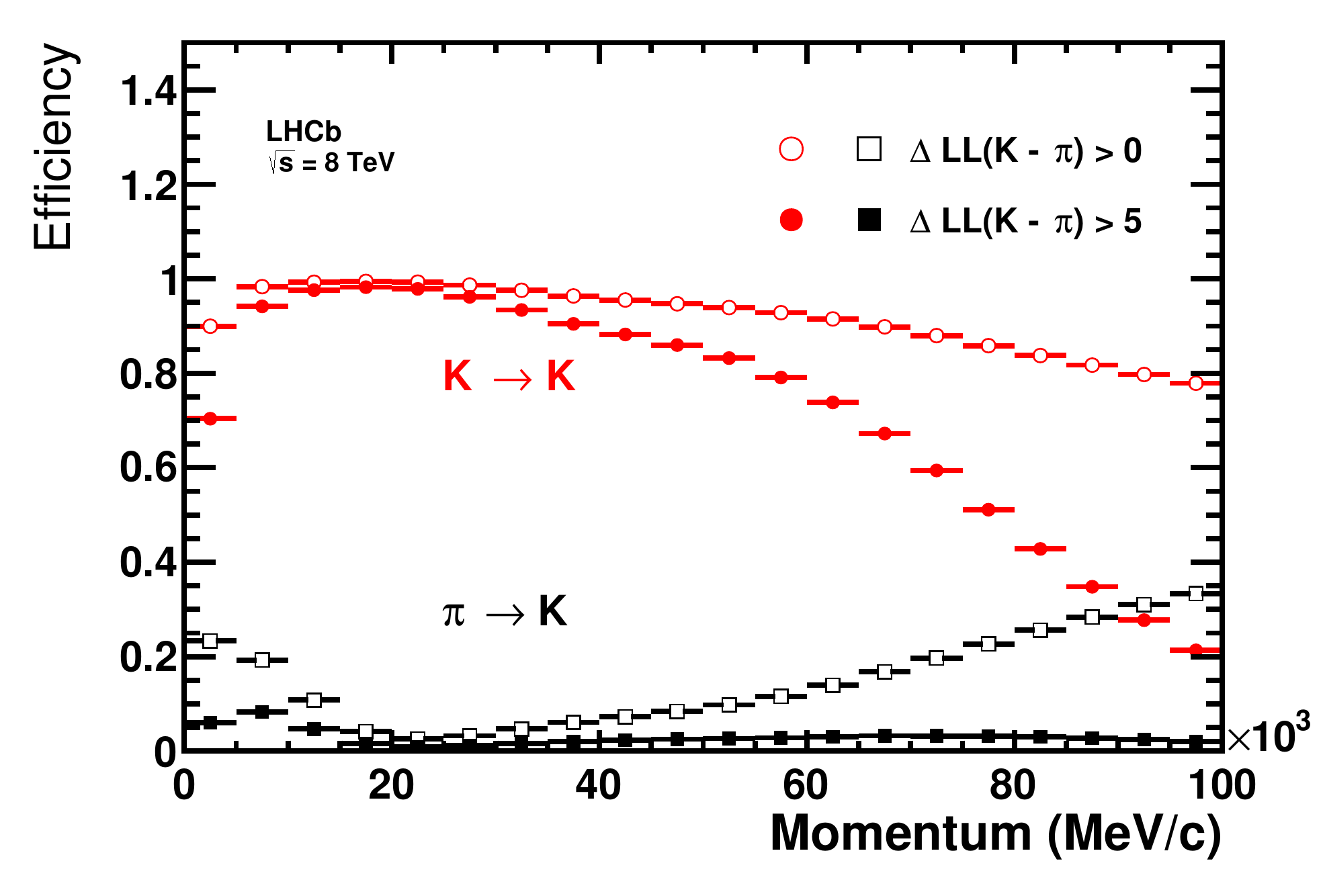}
\caption{Kaon ID performance (efficiency ($K \rightarrow K $ and mis-ID
of pions as Kaons ($\pi \rightarrow K$)) versus momentum for two 
different cuts in the likelihood difference (decision making variable). 
At the top the data correspond to 2015 and the bottom to the 
2012 period of data taking.}
\label{fig:PIDvsMom2015}
\end{figure}

Using particle samples of known species, as described above 
(section~\ref{calib}), it is possible to estimate the performance of the 
RICH system in terms of particle ID. The efficiency of the system 
can be traded against purity. Figure~\ref{fig:PIDvsMom2015} 
(top) shows the efficiency and mis-identification of kaons versus 
momentum for the 2015 data taking period.
Efficiency and mis-identification are shown for two different values of the 
decision-making variable (the difference in the likelihood). For a particular
physics study the performance can be tuned to match the requirements of
the analysis. The same data are shown at the bottom part of the figure for 
the 2012 data taking period (during Run I). The performance looks generally 
similar apart from the mis-identification curves at low momentum. It is 
clear that in 2015, after all the changes described in section~\ref{changes}, 
the performance is better. The purity has improved
without any sacrifice in efficiency.

\begin{figure*}
\centering
\includegraphics[width=\linewidth]{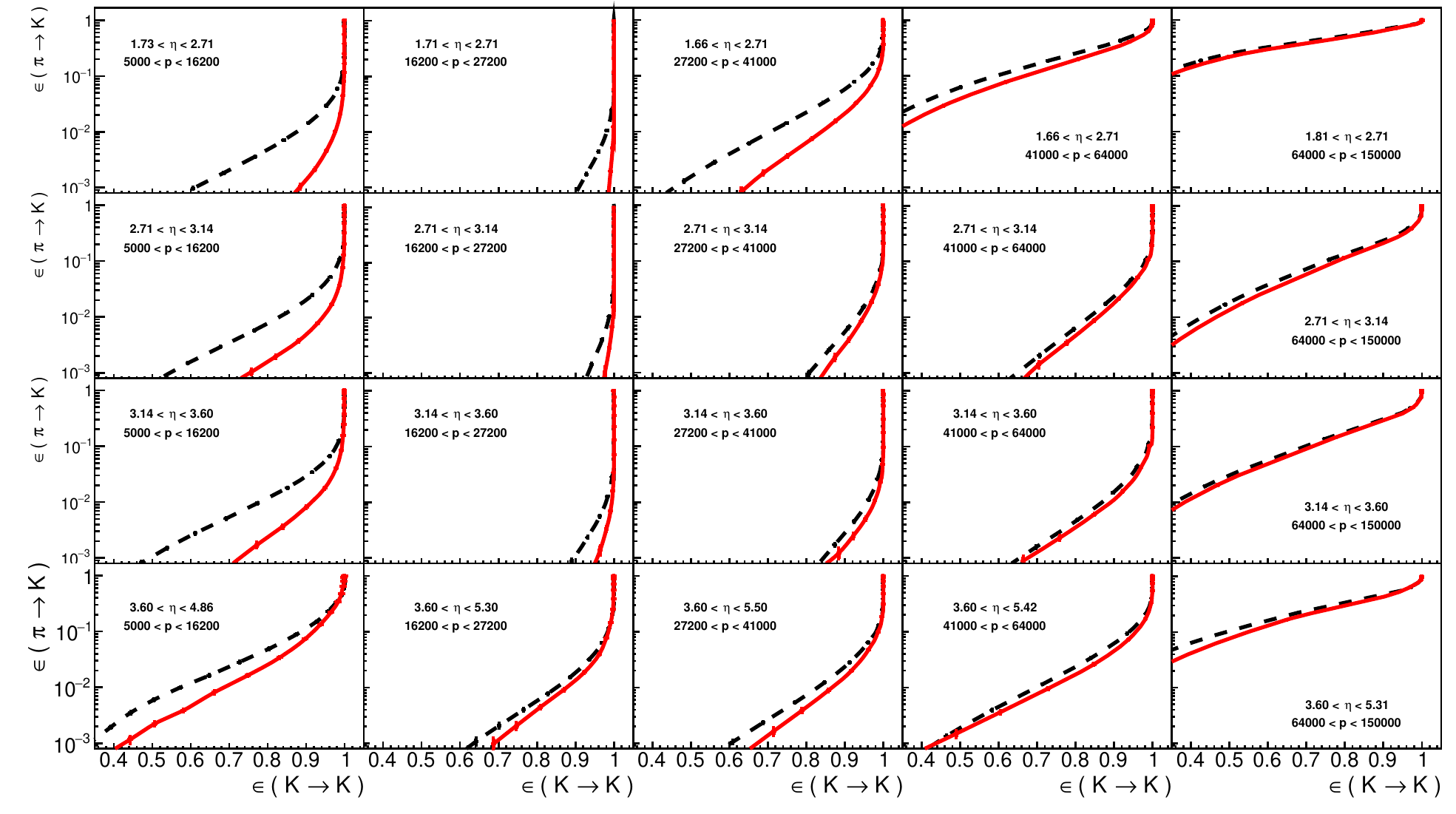}
\includegraphics[width=\linewidth]{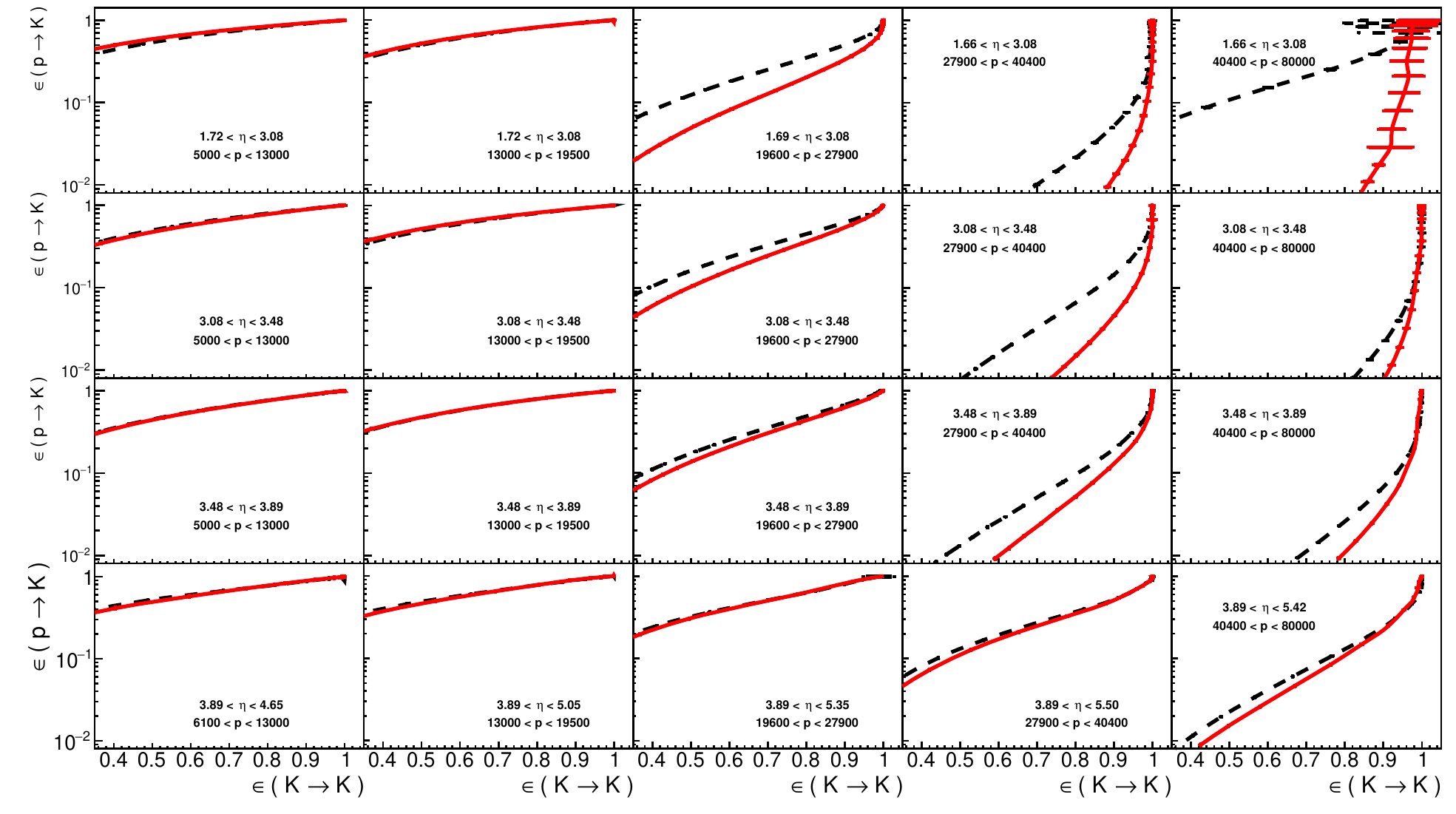}
\caption{Comparison of the performance of the RICH particle ID between 2012
(black dash lines) and 2015 (red solid lines). The top plot shows the
efficiency versus mis-identification for pions and kaons and the lower plot 
is the equivalent for kaons and protons.
The data have 
been re-weighted to correct for differences in the number of tracks in each
event and split in bins of momentum and pseudorapidity. 
Each point corresponds to a different value in the likelihood difference
between the two particles. It is clear from the 
comparison that the changes of the RICH system have had only positive
results. There is no area in the $p - \eta$ space where the performance
has deteriorated.}
\label{fig:perfComp}
\end{figure*}

Although Figure~\ref{fig:PIDvsMom2015} shows that the overall performance of 
the RICH system has improved in Run II, comparison with 2012 is not 
straightforward as the beam energy and data taking conditions are not 
the same. While studying the performance in 2015 it became clear that 
changing the trigger conditions and altering the mix of selected events
had an impact on the measured RICH performance. It is however also important 
to understand the intrinsic performance of the LHCb RICH system.
The parameters that most affect the performance of the RICH, apart 
from the particle momentum, are the complexity of the event (the number 
of tracks) and the pseudorapidity of the track.
The Run I data is re-weighted to correct for differences in the distribution 
of the number of tracks per event compared to Run II data.
After this step the performance is studied in bins of $\eta$ and momentum.
Figure~\ref{fig:perfComp} shows the comparison of the performance between
2012 (dash lines) and 2015 (solid lines). Each small 
plot shows the mis-identification versus the
efficiency for a range of momentum and $\eta$ for 2012 and 2015 data.
The top plot shows the performance in pion-kaon separation and the lower
plot the same for kaons and protons. 
The plots are produced by varying the difference in likelihood between the
two particles under test.
Looking at the performance of the RICH 
system in this way provides detailed and useful information. It is worth 
noting that the kaon threshold in RICH\,1 is ~9.4~GeV/$c$ and that
RICH\,2 covers angles $\eta \gtrsim 2.5$. In the comparison of the 2012-2015
performance it is clear that certain regions of phase space
have improved while in others the performance has not changed. 
In the entire $\eta - p$ phase-space
there is no performance degradation.
Any degradation in performance due to the removal of the aerogel would be 
seen at lower particle momenta, and particularly for $K$-proton separation. 
This is not seen, indicating that the benefit from the higher Cherenkov 
photon yield more than compensates for the loss of information from the 
aerogel.
It should be pointed out that 
the RICH particle ID is always used together with 
tracking and other subsystem information in the
event and the binary differentiation between two particle species 
represents the way it is actually used in physics studies.
 
% 
%%%%%%%%%%%%%%%%%%%%%%%%%%%%%%%%%%%%% 
% 

\section{Conclusions}
Extensive work took place following Run I in 2013-14 to prepare the LHCb 
RICH detectors for LHCb Run II.
The biggest change for the LHCb Run II was the 
removal of the aerogel radiator. A large number of photon detectors
was exchanged with refurbished ones having significantly better vacuum
quality and the readout electronics were better tuned for the 
LHCb data taking conditions. The re-circulation of the gas radiators was
also improved.

All of these changes have had a positive impact in the particle identification
performance. The separation between pions and kaons has been improved,
especially in the momentum region of 2--20~GeV/$c$.
Detailed studies and comparison between the 2012 and 2015 
show that this improvement is intrinsic to the RICH system and
not the result of any change in the data taking conditions.

%%%%%%%%%%%%%%%%%%%%%%%%%%%%%%%%%%%%%%%